\documentclass{article}

\usepackage[preprint]{neurips_2026}

\title{Tree-based Credit Assignment for Multi-Agent Memory System}

\usepackage{amsmath}
\usepackage{natbib}
\usepackage[utf8]{inputenc} 
\usepackage[T1]{fontenc}    

\usepackage{graphicx}
\usepackage{float}
\usepackage{subcaption}
\usepackage{wrapfig}
\usepackage{float}
\usepackage{url}            
\usepackage{booktabs}       
\usepackage{amsfonts}       
\usepackage{nicefrac}       
\usepackage{microtype}      
\usepackage{xcolor}         
\usepackage[colorlinks,
            linkcolor=red,
            anchorcolor=blue,
            citecolor=green
            ]{hyperref}
\usepackage{ragged2e} 
\usepackage{booktabs,makecell, multirow, tabularx}
\usepackage{colortbl}
\definecolor{TableRule}{HTML}{32383D}
\definecolor{TableGroupBlue}{HTML}{E6EEF8}
\definecolor{TableGroupGreen}{HTML}{E4F1EA}
\definecolor{TableGroupGold}{HTML}{F4ECD8}
\definecolor{TableGroupRose}{HTML}{F3E4E6}
\definecolor{TableAvgGray}{HTML}{EFEFEF}
\usepackage{algorithmic}
\usepackage[ruled,linesnumbered,vlined]{algorithm2e}

\usepackage{enumitem}
\usepackage{verbatim}
\usepackage{CJKutf8}

\newcommand{\ie}{\textit{i.e., }}
\newcommand{\eg}{\textit{e.g., }}

\definecolor{mygray}{rgb}{0.9, 0.9, 0.9}
\usepackage{tcolorbox} 
\usepackage{graphicx} 

\author{
Marina Mao\textsuperscript{1,2},\,
Alexandr Liu\textsuperscript{2}\,
Pengbo Li\textsuperscript{3},\,
Siheng Li\textsuperscript{4},\,
Bo Zhou\textsuperscript{2}$^*$,\,
Xiang Wang\textsuperscript{1}\,\thanks{Corresponding author. Email: chaysezhou@tencent.com, xiangwang@ustc.edu.cn.}\\
University of Science and Technology of China\\
\textsuperscript{2}LLM Department, Tencent\\
\textsuperscript{3}The Hong Kong University of Science and Technology\\
\textsuperscript{4}The Chinese University of Hong Kong\\
}

\begin{document}

\maketitle
\begin{abstract}
Memory systems are widely adopted to enhance LLMs for long-horizon tasks, and are commonly organized as multi-agent pipelines with memory building, summarizing, and retrieval agents. To empower this system, existing RL-based methods either apply final downstream task rewards (\eg QA accuracy) for all agents uniformly, which are coarse and ambiguous, or design task-specific rewards for agents on different subtasks, which require costly annotations (\eg key evidence) and are difficult to define reliably.
To address these limitations, we propose Tree-based Credit Assignment for Multi-Agent Memory Systems (TreeMem), which derives agent-specific credit from the final reward without task-specific annotations. Specifically, TreeMem extends the multi-agent pipeline (builder--summarizer--retrieval) into a tree structure, where each agent's outputs are expanded into multiple subsequent branches. The contribution of each agent is estimated via Monte Carlo averaging over its subsequent branches, capturing how intermediate agent actions may influence the final reward. This converts the coarse final reward into agent-specific optimization signals. These signals are then used to update all agent policies simultaneously, helping heterogeneous agents specialize effectively. Experiments on long-horizon benchmarks show that TreeMem improves memory system performance over strong baselines, validating the effectiveness of tree-structured credit assignment for the multi-agent memory system.
\end{abstract}

\section{Introduction}
\begin{figure}[t]
  \includegraphics[width=\columnwidth]{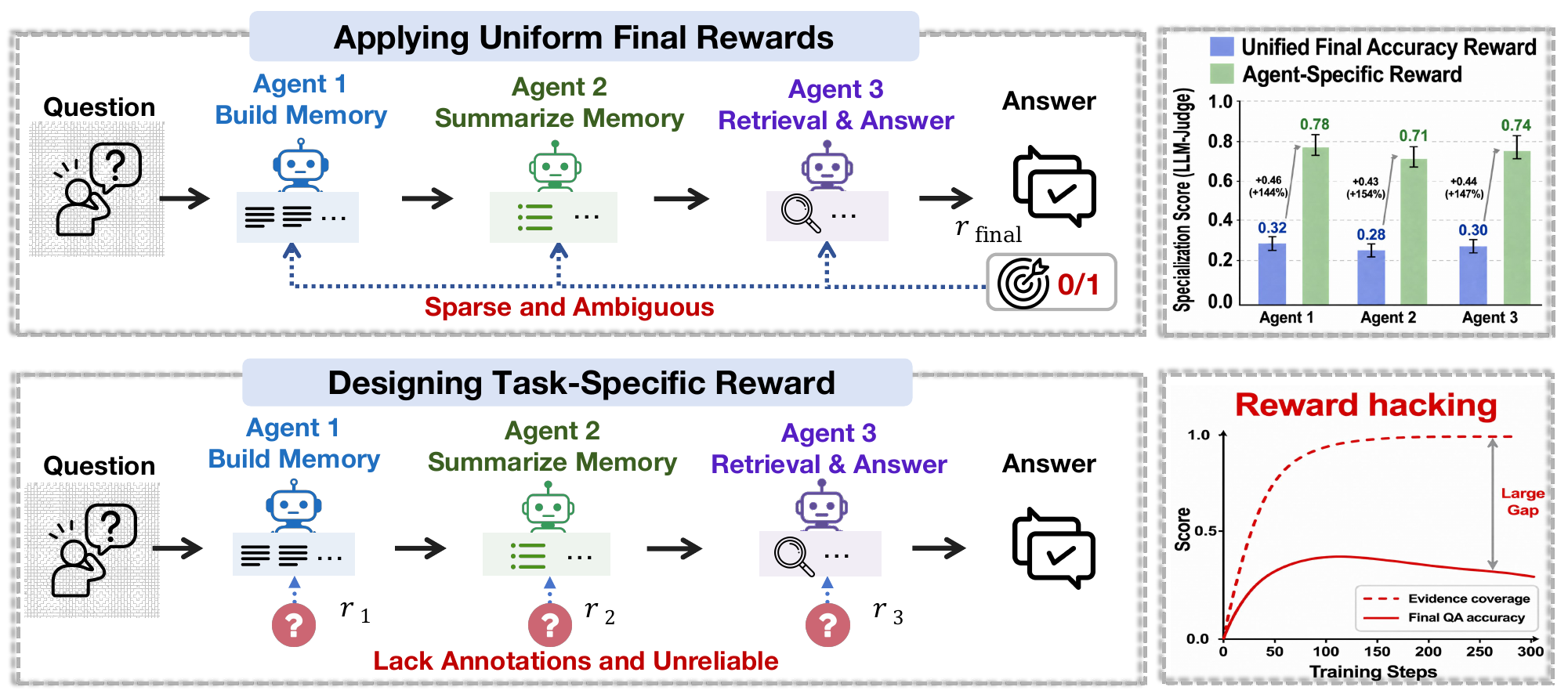}
  \caption{
 Demonstration and limitations of two training paradigms for existing RL-based methods in multi-agent memory systems.
  }
  \label{fig:teaser}
  \vspace{-6mm}
\end{figure}
Memory systems have recently emerged as an effective paradigm for extending large language models (LLMs) beyond their limited context windows \cite{hu2025memory,memorybank,HippoRAG, memoryt1, lightmem, assomem}, enabling models to store and dynamically incorporate relevant historical information for improved response quality in long-horizon QA tasks \cite{long-horizon,mem1} or reasoning tasks \cite{reasoningbank, remem}.
In practice, such systems are commonly implemented as multi-agent pipelines \cite{GAM, mirix, xu2025amem}, in which Builder agents first convert raw interaction histories into structured memory units \cite{mirix}, Summarizer agents abstract these units into higher-level representations \cite{memorybank}, and Retrieval agents select relevant information and answer downstream queries \cite{GAM,memagent, memgen}. To empower the multi-agent memory system, existing work has explored optimizing these agents with reinforcement learning (RL) \cite{mem-r1,mem1,memory-alpha,memagent}.

Existing RL-based methods optimize multi-agent memory pipelines primarily using two paradigms, as illustrated in Figure~\ref{fig:teaser}. The first applies the final downstream task rewards uniformly across all, such as final QA accuracy~\cite{mem-r1,mem1}. While this aligns training with the ultimate task objective, it provides coarse supervision and ambiguous signals for heterogeneous agents, limiting role-specific specialization. The second paradigm designs task-specific rewards for individual agents, such as evidence coverage for the Builder agent~\cite{CoMAM} or memory abstraction quality for the Summarizer agent~\cite{memory-alpha}, providing more direct intermediate supervision. However, such rewards require costly annotations and are difficult to define reliably, as memory is context-dependent and subjective; improperly crafted rewards may induce reward hacking~\cite{reward_hacking}, where agents optimize the manually defined reward without improving and possibly harming downstream task performance.

To address this limitation, we propose \emph{Tree-based Credit Assignment for Multi-Agent Memory Systems} (TreeMem), an RL training framework that derives agent-specific credit directly from the final downstream reward, without task-specific annotations. Specifically, we explicitly model the multi-agent memory pipeline as a tree structure: building upon existing multi-agent memory pipelines (builder--summarizer--retrieval), TreeMem expands each agent's sampled outputs into multiple subsequent branches, forming a tree of candidate trajectories that captures how intermediate agent actions may influence the final reward. The contribution of each intermediate agent is estimated via Monte Carlo averaging over its subsequent branches, a process that converts the coarse final reward into more granular, agent-specific optimization signals. These signals are then used to update all agent policies simultaneously, enabling heterogeneous memory agents to specialize effectively.
We evaluate TreeMem on long-horizon memory benchmarks and demonstrate that it consistently improves downstream task performance over strong baselines. Further analysis shows that TreeMem effectively propagates informative training signals to specific agents, highlighting the importance of fine-grained credit assignment in complex multi-agent LLM memory systems.

\section{Related Work}
\label{sec:related_work}

\paragraph{Multi-agent Memory Systems.}
Memory systems \cite{hu2025memory} for LLMs are increasingly organized as multi-agent pipelines \cite{mirix,GAM,xu2025amem,memorybank}, where specialized agents store, summarize, and retrieve conversational information for long-horizon tasks. Beyond prompt-based designs that specify the roles of memory agents, recent work has explored RL-based optimization to further enhance agents' capabilities in memory systems. One line of work trains memory agents with unified outcome rewards from downstream tasks. For example, memory-R1 \cite{mem-r1} and MEM1 \cite{mem1} use final QA accuracy rewards to improve memory management and memory retrieval agents. Such rewards are naturally aligned with the end objective, but they are sparse and ambiguous for different agents. Another line of work designs task-specific rewards for individual agents. Mem-$\alpha$ \cite{memory-alpha} and CoMAM \cite{CoMAM}, for instance, introduce rewards related to evidence coverage or memory quality to provide more specialized supervision. However, they require additional annotation or handcrafted criteria and may introduce reward hacking. Motivated by these limitations, we study how to derive agent-specific credit for multi-agent memory systems without relying on task-specific annotation.

\paragraph{Credit Assignment in LLM-based Multi-Agent Optimization.}
Multi-agent joint training has gained increasing attention recently \cite{comas, marshal}.  
Credit assignment is a central problem in multi-agent reinforcement learning, where the goal is to attribute delayed final results to each agent \cite{counterfactual_credit_assignment,cocoa_credit_assignment, marti, drmas, malt}.
Existing LLM-based multi-agent optimization methods commonly assign credit by applying the same final task reward to all agents, or by optimizing agents under a shared system-level objective \cite{naive_assign,hong2025multi,MAPORL,MARFT, rema}. This strategy is simple and directly aligned with downstream success, but it is also coarse: LLM agents are often differentiated by roles, prompts, or policies, and a uniform reward cannot indicate which agent should improve or specialize. Some recent work seeks finer-grained credit assignment in LLM reasoning by estimating the utility of intermediate steps through tree-branching rollouts \cite{TreePO,TreeGRPO, treerl} or progress-based signals \cite{agentPRM}. However, these methods mainly target the optimization of a single policy along one reasoning trajectory. In multi-agent memory systems, agent contributions are coupled through complex interactions, making it difficult to isolate the effect of each agent on the final outcome. As a result, credit assignment for specialization among heterogeneous memory agents remains underexplored. 
\section{Method}
\label{sec:method}
\begin{figure}[t]
  \includegraphics[width=\columnwidth]{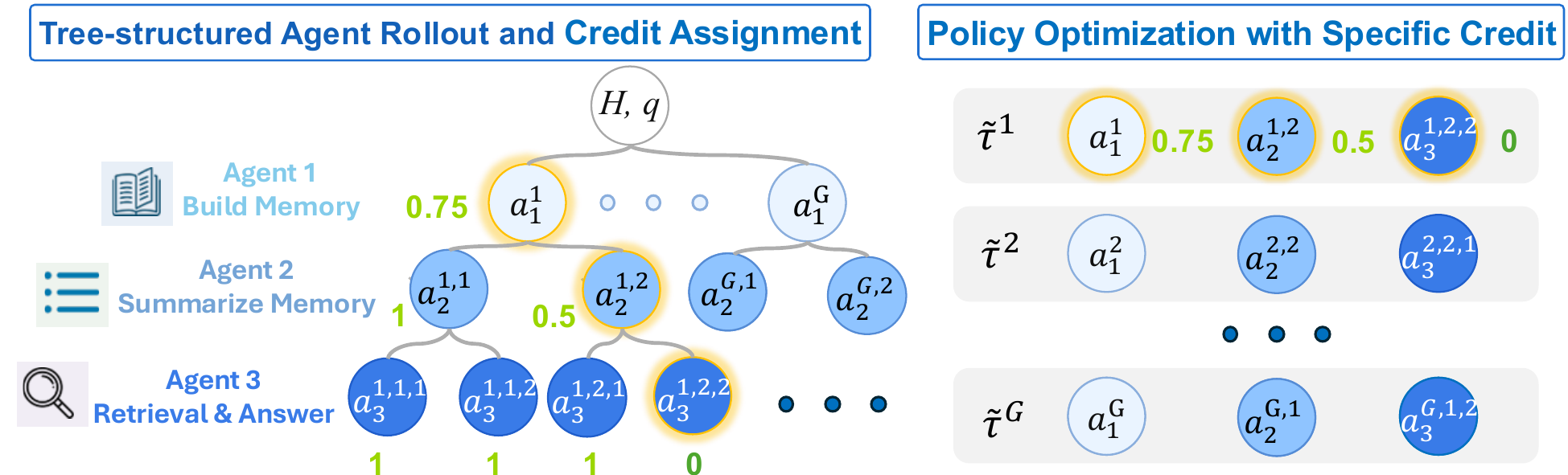}
  \vspace{2mm}
  \caption{
 Illustration of our proposed TreeMem: a tree-based multi-agent RL training framework for memory systems, expanding each agent's sampled actions into multiple branches to estimate agent-specific credit and specialize each agent's policy.
  }
  \label{fig:framework}

\end{figure}

In this section, we present TreeMem, a tree-based multi-agent RL training framework that derives agent-specific credit without task-specific annotations, as illustrated in Figure \ref{fig:framework}. We first formulate the multi-agent memory system and the final-reward optimization objective in Sec. \ref{sec:method:formulation}. We then describe tree-structured rollout and credit assignment in Sec. \ref{sec:method:tree_rollout}, which expands each agent's sampled actions into multiple subsequent branches to capture how intermediate actions influence the final outcome. Finally, we describe the corresponding policy optimization in Sec. \ref{sec:method:policy_optimization}, showing how the resulting signals promote specialization among memory agents.

\subsection{Formulation of Multi-Agent Memory Optimization}
\label{sec:method:formulation}

Let $H$ denote the long-horizon interaction history, $q$ the downstream query, and $p^*$ the ground-truth response to $q$. Motivated by existing multi-agent memory systems \cite{GAM, mem-r1, memorybank, CoMAM}, we instantiate a representative multi-agent memory workflow with a builder, a summarizer, and a retrieval-based responder. We denote these agent policies by $\Pi=\{\pi_{\theta}^{n}\}_{n=1}^{3}$, where the superscript identifies the agent. In the multi-agent workflow, the builder produces built fact-level memory $a_1$ from $H$, the summarizer produces summarized memory $a_2$ from $a_1$, and the retrieval-based responder produces the final answer $p$ from $q$ and the memories $(a_1,a_2)$:
\begin{equation}
a_1\sim\pi_{\theta}^{1}(\cdot\mid H),\qquad
a_2\sim\pi_{\theta}^{2}(\cdot\mid a_1),\qquad
a_3=p\sim\pi_{\theta}^{3}(\cdot\mid q,a_1,a_2).
\label{eq:agent-actions}
\end{equation}

For RL training, existing methods usually run the multi-agent pipeline to obtain a complete rollout $\tau=(a_1,a_2,a_3)$, and then compute the final downstream reward from the generated answer \cite{mem-r1,mem1}:
\begin{equation}
R(\tau)=F(p,p^*),
\label{eq:final-reward}
\end{equation}
where $p$ is the final answer produced by trajectory $\tau$, and $F$ is a scalar downstream evaluator. However, merely using the final reward to optimize all agents is sparse and ambiguous across different roles: a poor intermediate action may still be rewarded if later agents produce the correct answer, while a useful intermediate action may be penalized if later agents fail.
To provide more specific training signals for different agents, a natural solution is to design task-specific rewards for their respective tasks \cite{memory-alpha,CoMAM}, such as evidence coverage for the memory builder or memory quality for the summarizer. However, reliable task-specific annotations are costly to obtain, and handcrafted rewards may be noisy or misaligned with the final objective. TreeMem therefore focuses on fine-grained credit assignment for memory agents from final rewards. Specifically, we expand each agent's actions into multiple subsequent branches and estimate its specific effect on the final outcome by averaging the final rewards over the branches that depend on it. This procedure converts final-only feedback into agent-specific credit, which can then be used to optimize different policies and promote specialization among agents.

\subsection{Tree-Structured Rollout and Credit Assignment}
\label{sec:method:tree_rollout}

Given the formulation in Sec. \ref{sec:method:formulation}, TreeMem builds a tree-structured rollout during training by sampling multiple branches at each stage of the builder--summarizer--responder workflow, as illustrated on the left side of Figure \ref{fig:framework}. For an agent action, averaging the final rewards over its subsequent branches provides a more specific estimate of how that action affects the final outcome. Specifically, under the same history $H$, the memory builder samples $G$ candidate actions; for each builder action, the summarizer samples $J$ branches; and for each builder--summarizer pair, the retrieval-based responder samples $K$ final responses:
\begin{equation}
\begin{aligned}
a_1^{i} &\sim \pi_{\theta}^{1}(\cdot\mid H),
&& i=1,\dots,G,\\
a_2^{i,j} &\sim \pi_{\theta}^{2}(\cdot\mid a_1^{i}),
&& j=1,\dots,J,\\
a_3^{i,j,k}=p^{i,j,k} &\sim \pi_{\theta}^{3}(\cdot\mid q,a_1^{i},a_2^{i,j}),
&& k=1,\dots,K.
\end{aligned}
\label{eq:tree-branch-sampling}
\end{equation}
Here $a_1^{i}$ is the $i$-th built fact-level memory, $a_2^{i,j}$ is the $j$-th summarized memory conditioned on $a_1^{i}$, and $p^{i,j,k}$ is the $k$-th response generated from query $q$ and memories $(a_1^{i},a_2^{i,j})$. The parameter $G$ is the group size, \ie the number of memory builders' actions sampled for the same training example, while $J$ and $K$ are the branch sizes for the summarizer and the responder, respectively. Each builder action $a_1^i$ roots a subtree containing $J\cdot K$ complete trajectories.

This procedure produces $G\cdot J\cdot K$ complete trajectories, each corresponding to one path from a built fact-level memory to a final response. Since TreeMem does not assume intermediate supervision, rewards are computed only at the final leaf nodes:
\begin{equation}
\tau^{i,j,k}=(a_1^{i},a_2^{i,j},a_3^{i,j,k}),
\qquad
R^{i,j,k}=R(\tau^{i,j,k})=F(p^{i,j,k},p^*).
\label{eq:tree-trajectory-reward}
\end{equation}
Here $\tau^{i,j,k}$ denotes the complete trajectory formed by the $i$-th builder action, the $j$-th summarizer action under that builder action, and the $k$-th responder action under that builder--summarizer pair. $R^{i,j,k}$ is the scalar final reward assigned to this leaf trajectory, computed by applying the downstream evaluator $F$ to the generated response $p^{i,j,k}$ and the ground-truth response $p^*$. 

Given the tree-structured rollout, TreeMem assigns credit by propagating final rewards upward through Monte Carlo branch averaging. The final response already has an observed reward, while each upstream action is evaluated by averaging the final rewards over the sampled branches that depend on it. For the memory builder, we additionally apply a length penalty based on the output-to-input token ratio, which discourages copying the long history into memory verbatim. Specifically, the credits for the responder, summarizer, and builder actions are
\begin{equation}
\widehat{Q}_3^{i,j,k}=R^{i,j,k},\quad
\widehat{Q}_2^{i,j}=\frac{1}{K}\sum_{k=1}^{K}R^{i,j,k},\quad
\widehat{Q}_1^{i}=\frac{1}{JK}\sum_{j=1}^{J}\sum_{k=1}^{K}R^{i,j,k}+\frac{|a_1^i|}{|H|}.
\label{eq:agent-credit}
\end{equation}
Here $\widehat{Q}_3^{i,j,k}$ is the reward of a final response, $\widehat{Q}_2^{i,j}$ averages over the $K$ response branches following summarized memory $a_2^{i,j}$, and $\widehat{Q}_1^{i}$ averages over all summarization and response branches following built memory $a_1^{i}$, with a length penalty on the ratio between the length of fact-level memory $a_1^i$ and the input history $H$ to filter out redundancy. These estimates use the sampled subtree rooted at each action to convert final-only rewards into agent-specific credit.
Notably, the tree structure is used only during training to derive agent-specific credits for optimizing different agent policies. At deployment time, TreeMem performs inference with the optimized policies directly, allowing the memory system to invoke them in any order and at any frequency as needed, without executing tree expansion.

\subsection{Policy Optimization with Agent-Specific Credit}
\label{sec:method:policy_optimization}

After assigning credits over the full tree-structured rollout, TreeMem constructs a trajectory group by randomly selecting one complete trajectory from each of the $G$ subtrees rooted at the memory-builder actions $\{a_1^i\}_{i=1}^{G}$. This produces $G$ trajectories, one from each subtree, as illustrated on the right side of Figure~\ref{fig:framework}. For the subtree rooted at $a_1^i$, let $j_i\in\{1,\dots,J\}$ and $k_i\in\{1,\dots,K\}$ denote the selected summarizer and responder branch indices. The selected trajectory from the subtree rooted at $a_1^i$ and its corresponding agent-specific credits are:
\begin{equation}
\widetilde{\tau}^{i}=(a_1^{i},a_2^{i,j_i},a_3^{i,j_i,k_i}),\qquad
\widetilde{Q}_1^{i}=\widehat{Q}_1^{i},\qquad
\widetilde{Q}_2^{i}=\widehat{Q}_2^{i,j_i},\qquad
\widetilde{Q}_3^{i}=\widehat{Q}_3^{i,j_i,k_i},
\qquad i=1,\dots,G.
\label{eq:selected-trajectory-credit}
\end{equation}
This group sampling strategy enables compact yet informative policy optimization. Sampling one trajectory from each subtree preserves the diversity of memory-builder actions while avoiding optimization over the full rollout tree. Since the selected actions inherit credits computed from the full tree rollout, each agent is updated with agent-specific credit that reflects its expected downstream contribution, rather than a sparse final reward. For each agent $n$, the credits of its $G$ selected actions are normalized as advantages within the group:
\begin{equation}
\widehat{A}_n^{i}
=
\frac{
\widetilde{Q}_n^{i}-\operatorname{mean}_{r=1}^{G}\widetilde{Q}_n^{r}
}{
\operatorname{std}_{r=1}^{G}\widetilde{Q}_n^{r}+\epsilon
},
\qquad n\in\{1,2,3\}.
\label{eq:agent-advantage}
\end{equation}

TreeMem then optimizes the heterogeneous policies using the selected trajectory group and the normalized advantages in Eq. \ref{eq:agent-advantage}, following the group-relative principle of GRPO \cite{DeepSeekMath}. Unlike standard GRPO, which compares complete responses from a single policy, TreeMem computes the group score for each agent from its subtree-estimated credit $\widetilde{Q}_n^i$ to update different policies jointly. 

Let $c_n^i$ denote the input context of agent $n$ in trajectory $\widetilde{\tau}^{i}$, with $c_1^i=H$, $c_2^i=a_1^i$, and $c_3^i=(q,a_1^i,a_2^{i,j_i})$. Let $o_n^{i}$ denote the token sequence generated by agent $n$ under context $c_n^i$. For each agent, TreeMem optimizes the following clipped policy objective:
\begin{equation}
\mathcal{J}_n(\theta)
=
\frac{1}{G}
\sum_{i=1}^{G}
\frac{1}{|o_n^{i}|}
\sum_{t=1}^{|o_n^{i}|}
\min\left(
\rho_{n,i,t}\widehat{A}_n^{i},
\operatorname{clip}(\rho_{n,i,t},1-\epsilon_{\mathrm{clip}},1+\epsilon_{\mathrm{clip}})
\widehat{A}_n^{i}
\right),
\label{eq:optimization}
\end{equation}
where
$\rho_{n,i,t}
=
\frac{
\pi_{\theta}^{n}(o_{n,t}^{i}\mid c_n^i,o_{n,<t}^{i})
}{
\pi_{\theta_{\mathrm{old}}}^{n}(o_{n,t}^{i}\mid c_n^i,o_{n,<t}^{i})
}.$
Here $\theta_{\mathrm{old}}$ is the policy used to sample the trajectory group, and the clipping operation constrains the update from this sampling policy. The factor $1/|o_n^i|$ normalizes the objective by the action length, preventing longer generated actions from dominating the policy update.
Thus, TreeMem updates each policy using the subtree-estimated credit of its action, rather than assigning the final reward to all agents. This keeps optimization aligned with the final task objective while providing agent-specific training signals for the multi-agent memory pipeline.

\section{Experiments}
Our experiments are designed to examine TreeMem from five complementary perspectives:
\begin{itemize}[leftmargin=*, itemsep=0pt, parsep=0pt, topsep=0pt]
\item RQ1: How does TreeMem perform compared with representative baselines on long-horizon tasks?
\item RQ2: What's the impact of different RL-training signals on the memory system's performance?
\item RQ3: How does tree-based credit assignment affect different memory agents?
\item RQ4: How sensitive is TreeMem to the hyperparameters?
\item RQ5: What training and token overheads does TreeMem incur?
\end{itemize}

\begin{table*}[t]
\centering
\caption{Query-answering accuracy on PersonaMem under 32K, 128K, and 1M history lengths. \textbf{Bold} marks the overall best result, and \underline{underlining} marks the strongest compared method.}
\label{tab:main_results}
\setlength{\tabcolsep}{6pt}
\arrayrulecolor{TableRule}
\small
\begin{tabular}{l |cccc cccc}
\toprule
\multirow{2}{*}{\textbf{Method}} & \multicolumn{4}{c}{\cellcolor{TableGroupBlue}\textbf{Qwen}} & \multicolumn{4}{c}{\cellcolor{TableGroupGreen}\textbf{Llama}} \\
\cmidrule(lr){2-5}\cmidrule(lr){6-9}
& 32K & 128K & 1M & Average & 32K & 128K & 1M & Average \\
\midrule
\rowcolor{TableAvgGray}
& \multicolumn{8}{c}{\textbf{Context-based methods}}\\
\midrule
Base & 0.41 & 0.39 & 0.38 & \cellcolor{TableAvgGray}0.39 & 0.35 & 0.31 & 0.32 & \cellcolor{TableAvgGray}0.33\\
RAG \cite{RAG} & 0.48 & 0.45 & 0.41 & \cellcolor{TableAvgGray}0.45 & 0.43 & 0.39 & 0.36 & \cellcolor{TableAvgGray}0.39 \\
\midrule
\rowcolor{TableAvgGray}
& \multicolumn{8}{c}{\textbf{Prompt-driven memory}}\\
\midrule
CAM \cite{li2025cam} & 0.53 & 0.50 & 0.45 & \cellcolor{TableAvgGray}0.49 & 0.48 & 0.45 & 0.43 & \cellcolor{TableAvgGray}0.45\\
MemoryBank \cite{memorybank} & 0.51 & 0.47 & 0.44 & \cellcolor{TableAvgGray}0.47 & 0.45 & 0.42 & 0.39 & \cellcolor{TableAvgGray}0.42 \\
A-Mem \cite{xu2025amem} & 0.50 & 0.44 & 0.45 & \cellcolor{TableAvgGray}0.46 & 0.46 & 0.37 & 0.36 & \cellcolor{TableAvgGray}0.40 \\
Mem0 \cite{chhikara2025mem0} & 0.56 & 0.55 & 0.56 & \cellcolor{TableAvgGray}0.56 & 0.55 & 0.49 & 0.54 & \cellcolor{TableAvgGray}0.53\\
\midrule
\rowcolor{TableAvgGray}
& \multicolumn{8}{c}{\textbf{RL-optimized memory}}\\
\midrule
Mem1 \cite{mem1} & 0.59 & 0.57 & 0.58 & \cellcolor{TableAvgGray}0.58 & 0.56 & 0.58 & 0.61 & \cellcolor{TableAvgGray}0.58 \\
Mem-$\alpha$ \cite{memory-alpha} & 0.62 & 0.67 & 0.62 & \cellcolor{TableAvgGray}0.64 & 0.58 & 0.66 & 0.63 & \cellcolor{TableAvgGray}0.62\\
Memory-R1 \cite{mem-r1} & 0.58 & 0.60 & 0.60 & \cellcolor{TableAvgGray}0.59 & 0.57 & 0.61 & 0.60 & \cellcolor{TableAvgGray}0.59 \\
CoMAM \cite{CoMAM} & \underline{0.64} & \underline{0.70} & \underline{0.66} & \cellcolor{TableAvgGray}\underline{0.67} & \underline{0.62} & \underline{0.68} & \underline{0.69} & \cellcolor{TableAvgGray}\underline{0.66}\\
\midrule
Ours & \textbf{0.69} & \textbf{0.75} & \textbf{0.71} & \cellcolor{TableAvgGray}\textbf{0.72} & \textbf{0.64} & \textbf{0.74} & \textbf{0.72} & \cellcolor{TableAvgGray}\textbf{0.70} \\
\rowcolor{TableGroupRose}
\textbf{Impro.} & \textbf{+7.8\%} & \textbf{+7.1\%} & \textbf{+7.6\%} & \textbf{+7.5\%} & \textbf{+3.2\%} & \textbf{+8.8\%} & \textbf{+4.3\%} & \textbf{+5.5\%} \\
\bottomrule
\end{tabular}%
\arrayrulecolor{black}
\end{table*}

\begin{table*}[t]
\centering
\caption{More experimental results of TreeMem and baselines on the LongMemEval and LOCOMO benchmarks with Qwen2.5-7B backbone. \textbf{Bold} marks the overall best result, and \underline{underlining} marks the strongest compared method.}
\label{tab:longmemeval_locomo}
\setlength{\tabcolsep}{20pt}
\arrayrulecolor{TableRule}
\small
\begin{tabular}{l| ccc cc}
\toprule
\multirow{2}{*}{\textbf{Method}} & \multicolumn{3}{c}{\cellcolor{TableGroupBlue}\textbf{LongMemEval}} & \multicolumn{2}{c}{\cellcolor{TableGroupGreen}\textbf{LOCOMO}} \\
\cmidrule(lr){2-4}\cmidrule(lr){5-6}
& $F_1 \uparrow$ & $B_1 \uparrow$ & $J_1 \uparrow$ & $F_1 \uparrow$ & $B_1 \uparrow$ \\
\midrule
RAG & 18.27 & 14.57 & 22.20 & 8.97 & 7.27 \\
A-Mem & 41.55 & 36.58 & \underline{54.80} & 26.08 & 21.78 \\
Mem0 & 38.44 & 34.53 & 46.80 & 30.61 & 23.55 \\
GAM &51.75 & 48.96 & 52.00 & \underline{ 45.77} &  \underline{39.42} \\
Memory-R1& \underline{54.36}& \underline{50.02}& \underline{55.56} &43.14&36.14\\
\midrule
\rowcolor{TableGroupRose}
Ours &\textbf{63.46} & \textbf{52.61} & \textbf{58.25} & \textbf{48.88} & \textbf{43.38} \\
\bottomrule
\end{tabular}%
\arrayrulecolor{black}
\end{table*}

\subsection{Experimental Settings}
\label{sec: exp_settings}
\paragraph{Datasets.}
PersonaMem \cite{personamem} serves as our main testbed. It contains over 180 long conversational histories at three scales, 32K, 128K, and 1M tokens, together with roughly 6,000 questions from seven categories that require evidence from prior interactions. To test whether the observed behavior extends beyond this benchmark, we also report results on LongMemEval \cite{longmemeval} and LOCOMO \cite{LOCOMO}. Appendix~\ref{app: dataset} gives the full dataset description.

\paragraph{Baselines.}
We compare TreeMem against existing methods under three memory regimes.
\textit{Context-based methods} use the conversation history directly as the LLM context without converting it into explicit memories: Base conditions the LLM on the available conversation history within the limited context window, while RAG \cite{RAG} retrieves relevant raw history for response generation.
\textit{Prompt-driven memory systems} include CAM \cite{li2025cam}, MemoryBank \cite{memorybank}, A-Mem \cite{xu2025amem}, and Mem0 \cite{chhikara2025mem0}, which construct multi-agent memory systems through prompting or manually designed operations.
\textit{RL-optimized memory systems} include Mem1 \cite{mem1}, Mem-$\alpha$ \cite{memory-alpha}, Memory-R1 \cite{mem-r1}, and CoMAM \cite{CoMAM}, where reinforcement learning is used to optimize memory construction or retrieval.
Appendix \ref{app: baselines} provides concise descriptions of these baselines and their implementation details.

\paragraph{Implementation Details and Evaluation Metrics.}
Because PersonaMem's queries are multiple-choice, we report average query-answer selection accuracy for each history length (32K, 128K, and 1M). Following the official LongMemEval protocol, we report $F_1$, $B_1$, and $J_1$: $F_1$ measures token-level precision, $B_1$ denotes unigram BLEU, and $J_1$ is an LLM-as-judge score computed by GPT-4o. For LOCOMO, we report the overall $F_1$ and $B_1$ across different question types. To ensure fair comparisons, all methods are evaluated under identical experimental settings and with the same LLM backbone. For PersonaMem, we evaluate all methods with both Qwen \cite{bai2025qwen2} and Llama \cite{Llama}. The Builder agent, Summarizer agent, and reward model $\mathcal{V}$ are instantiated with Qwen2.5-3B-Instruct or Llama-3.2-3B-Instruct, while the Retrieval agent uses Qwen2.5-7B-Instruct or Llama-3.1-8B-Instruct. The context window is set to 32K tokens for all models. All results reported in the tables are averaged over at least three independent runs with different random seeds. Table~\ref{tab:main_results} reports results for both backbone families, whereas subsequent ablation studies use Qwen by default. Prompts, hyperparameters, and more implementation details are provided in Appendix \ref{app: implementation}.

\subsection{Main Results on Long-Horizon Memory Tasks (RQ1)}

To evaluate whether TreeMem improves long-horizon memory tasks over existing baselines, we conduct experiments on PersonaMem under three history-length settings (32K, 128K, and 1M) and two backbone families (Qwen and Llama), as reported in Table \ref{tab:main_results}. Compared with context-based methods, prompt-driven memory systems and RL-optimized memory systems benefit from converting raw dialogue histories into compact, reusable memories, thereby mitigating context-window limitations and reducing the burden on retrieval. RL-optimized memory systems further outperform prompt-driven memory systems, indicating the importance of optimizing memory agent policies with task feedback in multi-agent memory systems.
TreeMem achieves the best performance among leading RL-optimized memory systems, outperforming methods that rely on final-answer rewards (\eg Mem1 \cite{mem1} and Memory-R1 \cite{mem-r1}) as well as methods with manually designed task-specific rewards, such as memory-quality or evidence-coverage signals (\eg Mem-$\alpha$ \cite{memory-alpha} and CoMAM \cite{CoMAM}). This validates the effectiveness of tree-based credit assignment for multi-agent memory optimization: it maintains alignment with system-level performance while providing agent-specific training signals.
Results on LongMemEval and LOCOMO in Table \ref{tab:longmemeval_locomo} further show that TreeMem generalizes to benchmarks with different data distributions, achieving the best performance on all reported metrics.

\begin{table*}[t]
\centering
\caption{Performance comparison of multi-agent memory systems trained with different reward signals: final accuracy reward, task-specific rewards, and our agent-specific rewards.}
\label{tab: abla}
\setlength{\tabcolsep}{16pt}
\arrayrulecolor{TableRule}
\small
\begin{tabular}{l|ccc ccc}
\toprule
\multirow{2}{*}{\textbf{Method}} & \multicolumn{3}{c}{\cellcolor{TableGroupBlue}\textbf{Qwen}} & \multicolumn{3}{c}{\cellcolor{TableGroupGreen}\textbf{Llama}} \\
\cmidrule(lr){2-4}\cmidrule(lr){5-7}
& 32K & 128K & 1M & 32K & 128K & 1M \\
\midrule
No train & 0.50 & 0.46 & 0.44 & 0.52 & 0.47 & 0.43 \\
\midrule
+ $R_\text{final}$ & 0.60 & 0.64 & 0.60 & 0.61 & 0.65 & 0.67 \\
+ $R_\text{task}$ & 0.57 & 0.65 & 0.61 & 0.60 & 0.61 & 0.64 \\
+ $R_\text{final} + w R_\text{task}$ & 0.62 & 0.65 & 0.63 & 0.62 & 0.65 & 0.66 \\
\midrule
\rowcolor{TableGroupRose}
Ours & \textbf{0.69} & \textbf{0.75} & \textbf{0.71} & \textbf{0.64} & \textbf{0.74} & \textbf{0.72} \\
\bottomrule
\end{tabular}%
\arrayrulecolor{black}
\end{table*}

\begin{table*}[t]
\centering
\caption{Ablation study on tree-based credit assignment across different agents. w/o tree trains all agents without tree-based credit assignment; Builder w/o tree and Summarizer w/o tree remove it only from the Builder and Summarizer agents, respectively.}
\label{tab: abla_tree}
\setlength{\tabcolsep}{15pt}
\arrayrulecolor{TableRule}
\small
\begin{tabular}{l|ccc ccc}
\toprule
\multirow{2}{*}{\textbf{Method}} & \multicolumn{3}{c}{\cellcolor{TableGroupBlue}\textbf{Qwen}} & \multicolumn{3}{c}{\cellcolor{TableGroupGreen}\textbf{Llama}} \\
\cmidrule(lr){2-4}\cmidrule(lr){5-7}
& 32K & 128K & 1M & 32K & 128K & 1M \\
\midrule
No train & 0.50 & 0.46 & 0.44 & 0.52 & 0.47 & 0.43 \\
\midrule
w/o tree & 0.60 & 0.64 & 0.60 & 0.61 & 0.65 & 0.67 \\
Builder w/o tree & 0.65 & 0.65 & 0.60 & 0.62 & 0.67 & 0.68 \\
Summarizer w/o tree & 0.67 & 0.62 & 0.61 & 0.62 & 0.71 & 0.70 \\
\midrule
\rowcolor{TableGroupRose}
Ours & \textbf{0.69} & \textbf{0.75} & \textbf{0.71} & \textbf{0.64} & \textbf{0.74} & \textbf{0.72} \\
\bottomrule
\end{tabular}%
\arrayrulecolor{black}
\end{table*}

\subsection{Effect of Different Reward Signals (RQ2)}

To investigate how RL training signals affect the performance of memory systems, we compare several training variants in Table \ref{tab: abla}. Here, $R_\text{final}$ denotes training all agents uniformly using the final QA accuracy reward; $R_\text{task}$ refers to designing task-specific rewards for evidence coverage, memory quality, and retrieval precision tailored to different memory agents; and $R_\text{final} + wR_\text{task}$ integrates the two reward formulations. All RL-trained variants outperform the ``no train'' baseline, verifying that reinforcement learning yields effective optimization for multi-agent memory systems.
However, relying solely on the final reward delivers only sparse system-level supervision, which limits the specialization of individual agents. Task-specific rewards provide finer-grained guidance, though manually designed rewards are often unreliable and may not closely match downstream QA performance. Combining these two reward types yields a stronger training signal, yet TreeMem still achieves the best performance across all settings. This implies that straightforward fusion of the two reward components is inherently difficult to balance. By contrast, mapping high-level downstream rewards to dedicated credit signals for each agent can better align with overall system performance while enabling effective agent specialization.

\subsection{Impact of Tree-Based Credit Assignment on Each Agent (RQ3)}
\begin{figure}[t]
  \includegraphics[width=\columnwidth]{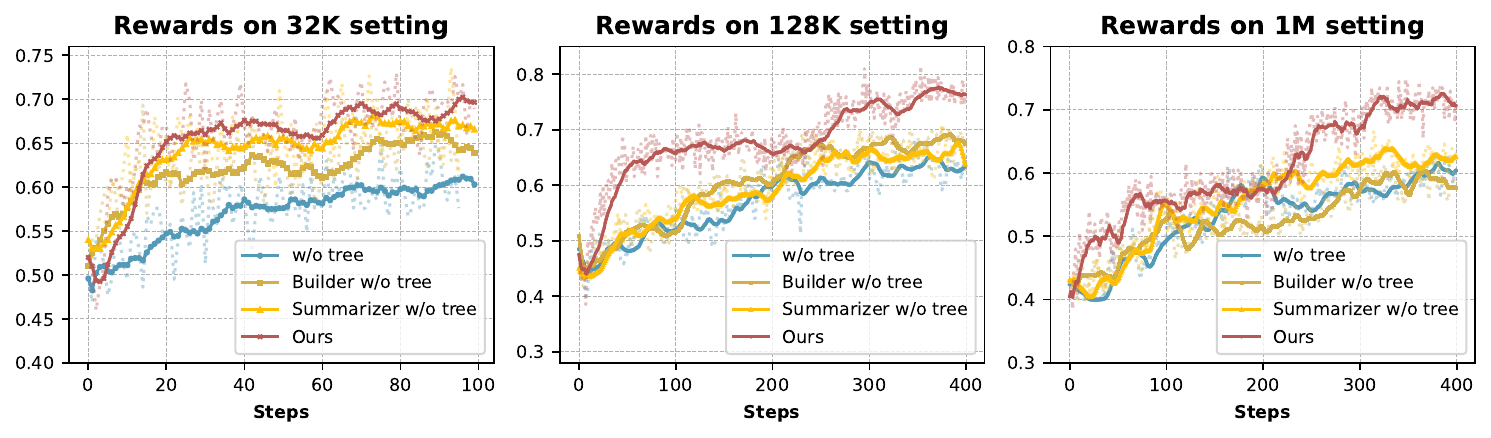}
  \caption{
Reward curve comparison of applying tree-based specific rewards to each agent individually.
  }
  \label{fig: exp_rewards}

\end{figure}

To evaluate how tree-based credit assignment affects different memory agents, we conduct an agent-level ablation study of the rollout-tree signal, with final performance reported in Table \ref{tab: abla_tree} and training rewards visualized in Figure \ref{fig: exp_rewards}. The variants include ``No train'' where no reinforcement learning is applied to any agent, ``w/o tree'' where all agents are trained uniformly with final accuracy rewards, ``Builder w/o tree'' and ``Summarizer w/o tree'' where the tree-based signal is removed only from the Builder or Summarizer agent, respectively, and ``Ours'' where tree-based credit assignment is enabled for all agents. As shown in the table and figure, across all context lengths, our TreeMem with tree-based credit assignment enabled for all agents consistently achieves the best final performance, while variants without tree-based credit perform the worst. Even when the agent-specific signal is removed from only one agent, either the Builder or the Summarizer, the corresponding variants ``Builder w/o tree'' and ``Summarizer w/o tree'' underperform our TreeMem, with their reward curves failing to reach the same performance ceiling. This pattern confirms that tree-based credit benefits every agent, as each agent relies on agent-specific credit to specialize its role in the memory system, which in turn enables the memory system to achieve better performance. Additionally, to further verify that tree-based agent-specific credit facilitates the specialization of individual memory agents, we adopt LLM-Judge (GPT-4o) scoring to quantitatively evaluate the quality of agent-generated outputs. As illustrated in the left and right subplots of Figure \ref{fig:teaser}, agents optimized with our agent-specific reward mechanism attain higher specialization scores than those trained solely with final accuracy rewards. We further provide qualitative generation examples in Appendix \ref{app:agent_specialization} to intuitively demonstrate such agent specialization.

\subsection{Sensitivity to Hyperparameters (RQ4)}
\begin{figure}[t]
  \includegraphics[width=\columnwidth]{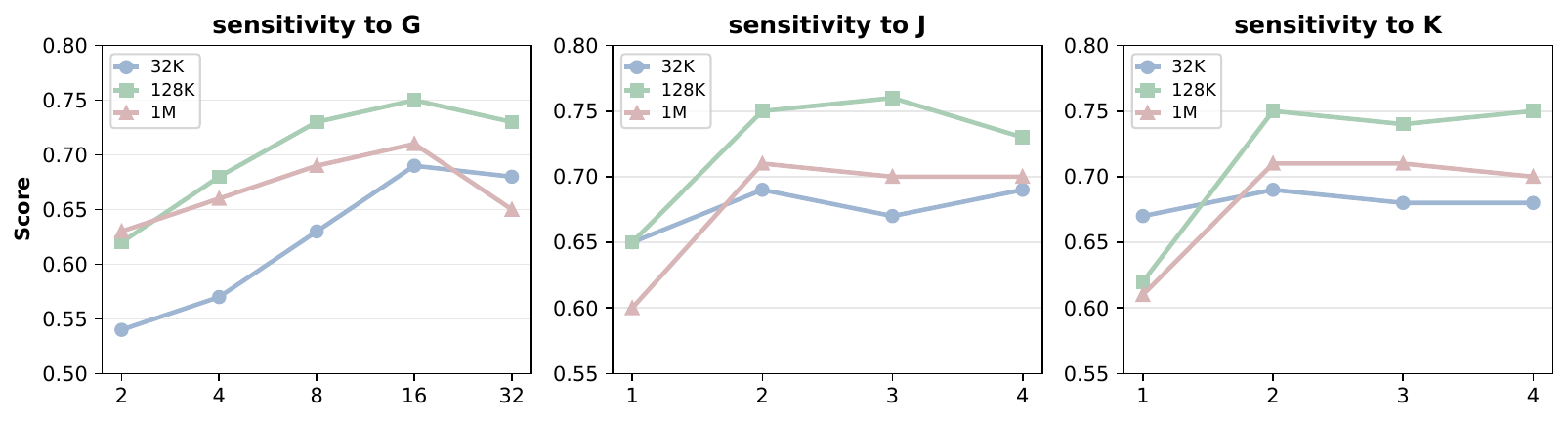}
  \caption{
Sensitivity analysis of TreeMem with respect to the hyperparameters group size $G$, branch size $J$, and branch size $K$, as defined in Eq. \eqref{eq:tree-branch-sampling}.
  }
  \label{fig: exp_sensitivity}

\end{figure}
To evaluate TreeMem’s sensitivity to its core hyperparameters, we vary the Builder group size $G$ from 2 to 32, the Summarizer branch count $J$ from 1 to 4, and the Retrieval branch count $K$ from 1 to 4. The group size $G$ controls the training efficiency and stability of GRPO optimization, while the branch counts $J$ and $K$ trade off credit-estimation quality and rollout computational cost.
As illustrated in Figure \ref{fig: exp_sensitivity}, all configurations exhibit clear performance trends across different context lengths. Small values of $J$ and $K$ make credit estimates more susceptible to sampling noise and result in lower scores. As $J$ and $K$ increase, performance continuously improves, yet the marginal gain gradually diminishes; adding extra branches brings no further performance improvement once a sufficient number is reached. In comparison, the Builder group size $G$ is designed exclusively for GRPO training. Its performance rises steadily as $G$ grows and eventually stabilizes at a moderate value.
Overall, TreeMem retains stable and robust performance over a wide range of hyperparameter choices. It works well without exhaustive grid search or meticulous parameter tuning, demonstrating strong hyperparameter insensitivity and practical applicability of our framework.

\subsection{Efficiency of TreeMem (RQ5)}
\begin{figure}[t]
  \includegraphics[width=\columnwidth]{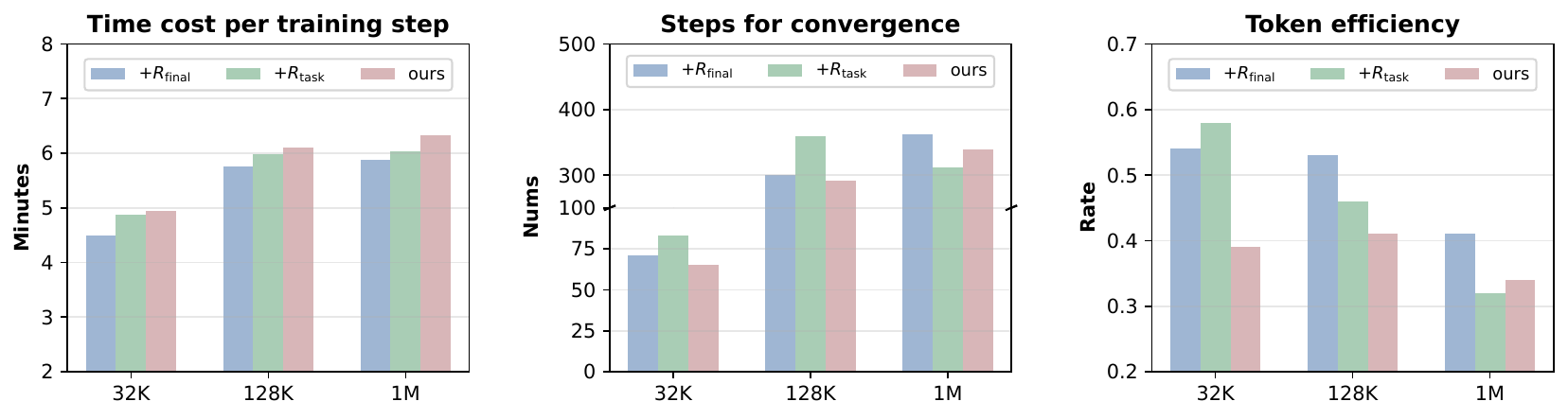}
  \caption{
The efficiency analysis of our TreeMem against two training baselines: ``+$R_\text{final}$''
 (final QA accuracy rewards) and ``+ $R_\text{task}$''
 (task-specific rewards). Across context lengths (32K, 128K, 1M), the left plot compares per-step training time, the middle plot compares steps to convergence, and the right plot evaluates token efficiency as the ratio of memory tokens (from the memory builder/summarizer) to original conversation history tokens, where a lower rate means higher efficiency.
  }
  \label{fig: exp_efficiency}
\end{figure}

To quantify TreeMem’s efficiency, we analyze training and token efficiency in Figure \ref{fig: exp_efficiency}. At training time, TreeMem’s tree-structured rollout introduces additional computation due to its branching factors, which we mitigate via sampling rollouts during policy optimization as introduced in Sec. \ref{sec:method:policy_optimization}, resulting in only marginal per-step overhead relative to the ``+$R_\text{final}$'' and ``+ $R_\text{task}$'' variants. Our agent-specific reward design also accelerates convergence, requiring fewer training steps to stabilize. For token efficiency (measured as the ratio of constructed memory tokens to original conversation tokens, with lower values indicating higher efficiency), TreeMem outperforms both baselines across all context lengths by enabling agent specialization and applying length penalties to compress long histories. At inference, TreeMem adopts the same memory architecture and agent backbone as baseline methods. It improves token efficiency to lower per-invocation computational overhead, while requiring no extra rollout branches. Overall, TreeMem introduces negligible extra computational burden during both training and inference, demonstrating superior computational efficiency compared with baseline alternatives.
\section{Conclusion}

In this work, we address a key limitation of existing training paradigms for multi-agent memory systems: assigning final downstream rewards uniformly to all memory agents is overly coarse to guide heterogeneous agents, while designing agent-specific task rewards demands costly manual annotations and lacks reliability. Specifically, we present TreeMem, a tree-based credit assignment framework for RL optimization of multi-agent memory systems. By extending the multi-agent pipeline to a tree structure and adopting Monte Carlo averaging over branch outputs, TreeMem accurately estimates the individual contribution of each agent. Thus, it could transform coarse final downstream accuracy rewards into fine-grained, agent-specific learning signals for different agents, enabling agent specialization and boosting overall memory system performance.
Extensive evaluations on long-horizon benchmarks verify the efficacy of TreeMem, which consistently outperforms leading memory system baselines across diverse experimental settings.

\bibliographystyle{unsrt}
\bibliography{references}

\begin{thebibliography}{10}

\bibitem{hu2025memory}
Yuyang Hu, Shichun Liu, Yanwei Yue, Guibin Zhang, Boyang Liu, Fangyi Zhu, Jiahang Lin, Honglin Guo, Shihan Dou, Zhiheng Xi, et~al.
\newblock Memory in the age of ai agents.
\newblock {\em arXiv preprint arXiv:2512.13564}, 2025.

\bibitem{memorybank}
Wanjun Zhong, Lianghong Guo, Qiqi Gao, He~Ye, and Yanlin Wang.
\newblock Memorybank: Enhancing large language models with long-term memory.
\newblock In {\em {AAAI}}, pages 19724--19731. {AAAI} Press, 2024.

\bibitem{HippoRAG}
Bernal~Jimenez Gutierrez, Yiheng Shu, Yu~Gu, Michihiro Yasunaga, and Yu~Su.
\newblock Hipporag: Neurobiologically inspired long-term memory for large language models.
\newblock In {\em NeurIPS}, 2024.

\bibitem{memoryt1}
Yiming Du, Baojun Wang, Yifan Xiang, Zhaowei Wang, Wenyu Huang, Boyang XUE, Bin Liang, Xingshan Zeng, Fei Mi, Haoli Bai, Lifeng Shang, Jeff~Z. Pan, Yuxin Jiang, and Kam-Fai Wong.
\newblock Memory-t1: Reinforcement learning for temporal reasoning in multi-session agents.
\newblock In {\em ICLR}, 2026.

\bibitem{lightmem}
Jizhan Fang, Xinle Deng, Haoming Xu, Ziyan Jiang, Yuqi Tang, Ziwen Xu, Shumin Deng, Yunzhi Yao, Mengru Wang, Shuofei Qiao, Huajun Chen, and Ningyu Zhang.
\newblock Lightmem: Lightweight and efficient memory-augmented generation.
\newblock In {\em ICLR}, 2026.

\bibitem{assomem}
Kai Zhang, Xinyuan Zhang, Ejaz Ahmed, Hongda Jiang, Caleb Kumar, Kai Sun, Zhaojiang Lin, Sanat Sharma, Shereen Oraby, AARON COLAK, Ahmed~A Aly, Anuj Kumar, Xiaozhong Liu, and Xin~Luna Dong.
\newblock Assomem: Scalable memory {QA} with multi-signal associative retrieval.
\newblock In {\em ICLR}, 2026.

\bibitem{long-horizon}
Mengkang Hu, Tianxing Chen, Qiguang Chen, Yao Mu, Wenqi Shao, and Ping Luo.
\newblock Hiagent: Hierarchical working memory management for solving long-horizon agent tasks with large language model.
\newblock In {\em {ACL} {(1)}}, pages 32779--32798. Association for Computational Linguistics, 2025.

\bibitem{mem1}
Zijian Zhou, Ao~Qu, Zhaoxuan Wu, Sunghwan Kim, Alok Prakash, Daniela Rus, Jinhua Zhao, Bryan Kian~Hsiang Low, and Paul~Pu Liang.
\newblock {MEM}1: Learning to synergize memory and reasoning for efficient long-horizon agents.
\newblock In {\em ICLR}, 2026.

\bibitem{reasoningbank}
Siru Ouyang, Jun Yan, I-Hung Hsu, Yanfei Chen, Ke~Jiang, Zifeng Wang, Rujun Han, Long Le, Samira Daruki, Xiangru Tang, Vishy Tirumalashetty, George Lee, Mahsan Rofouei, Hangfei Lin, Jiawei Han, Chen-Yu Lee, and Tomas Pfister.
\newblock Reasoningbank: Scaling agent self-evolving with reasoning memory.
\newblock In {\em ICLR}, 2026.

\bibitem{remem}
Yiheng Shu, Saisri~Padmaja Jonnalagedda, Xiang Gao, Bernal~Jim{\'e}nez Guti{\'e}rrez, Weijian Qi, Kamalika Das, Huan Sun, and Yu~Su.
\newblock {REM}em: Reasoning with episodic memory in language agent.
\newblock In {\em ICLR}, 2026.

\bibitem{GAM}
BY~Yan, Chaofan Li, Hongjin Qian, Shuqi Lu, and Zheng Liu.
\newblock General agentic memory via deep research.
\newblock {\em arXiv preprint arXiv:2511.18423}, 2025.

\bibitem{mirix}
Yu~Wang and Xi~Chen.
\newblock {MIRIX:} multi-agent memory system for llm-based agents.
\newblock {\em CoRR}, abs/2507.07957, 2025.

\bibitem{xu2025amem}
Wujiang Xu, Zujie Liang, Kai Mei, Hang Gao, Juntao Tan, and Yongfeng Zhang.
\newblock A-mem: Agentic memory for {LLM} agents.
\newblock In {\em NeurIPS}, 2025.

\bibitem{memagent}
Hongli Yu, Tinghong Chen, Jiangtao Feng, Jiangjie Chen, Weinan Dai, Qiying Yu, Ya{-}Qin Zhang, Wei{-}Ying Ma, Jingjing Liu, Mingxuan Wang, and Hao Zhou.
\newblock Memagent: Reshaping long-context {LLM} with multi-conv {RL}-based memory agent.
\newblock In {\em ICLR}, 2026.

\bibitem{memgen}
Guibin Zhang, Muxin Fu, and Shuicheng YAN.
\newblock Memgen: Weaving generative latent memory for self-evolving agents.
\newblock In {\em ICLR}, 2026.

\bibitem{mem-r1}
Sikuan Yan, Xiufeng Yang, Zuchao Huang, Ercong Nie, Zifeng Ding, Zonggen Li, Xiaowen Ma, Hinrich Sch{\"{u}}tze, Volker Tresp, and Yunpu Ma.
\newblock Memory-r1: Enhancing large language model agents to manage and utilize memories via reinforcement learning.
\newblock {\em CoRR}, abs/2508.19828, 2025.

\bibitem{memory-alpha}
Yu~Wang, Ryuichi Takanobu, Zhiqi Liang, Yuzhen Mao, Yuanzhe Hu, Julian McAuley, and Xiaojian Wu.
\newblock Mem-$\{$$\backslash$alpha$\}$: Learning memory construction via reinforcement learning.
\newblock {\em arXiv preprint arXiv:2509.25911}, 2025.

\bibitem{CoMAM}
Wenyu Mao, Haoyang Liu, Zhao Liu, Haosong Tan, Yaorui Shi, Jiancan Wu, An~Zhang, and Xiang Wang.
\newblock Collaborative multi-agent optimization for personalized memory system.
\newblock {\em CoRR}, abs/2603.12631, 2026.

\bibitem{reward_hacking}
Joar Skalse, Nikolaus H.~R. Howe, Dmitrii Krasheninnikov, and David Krueger.
\newblock Defining and characterizing reward gaming.
\newblock In {\em NeurIPS}, 2022.

\bibitem{comas}
Xiangyuan Xue, Yifan Zhou, Guibin Zhang, Zaibin Zhang, Yijiang Li, Chen Zhang, Zhenfei Yin, Philip Torr, Wanli Ouyang, and LEI BAI.
\newblock Co{MAS}: Co-evolving multi-agent systems via interaction rewards.
\newblock In {\em ICLR}, 2026.

\bibitem{marshal}
Huining Yuan, Zelai Xu, Zheyue Tan, Xiangmin Yi, Mo~Guang, Kaiwen Long, Haojia Hui, Boxun Li, Xinlei Chen, Bo~Zhao, Xiao-Ping Zhang, Chao Yu, and Yu~Wang.
\newblock {MARSHAL}: Incentivizing multi-agent reasoning via self-play with strategic {LLM}s.
\newblock In {\em ICLR}, 2026.

\bibitem{counterfactual_credit_assignment}
Thomas Mesnard, Theophane Weber, Fabio Viola, Shantanu Thakoor, Alaa Saade, Anna Harutyunyan, Will Dabney, Thomas~S. Stepleton, Nicolas Heess, Arthur Guez, Eric Moulines, Marcus Hutter, Lars Buesing, and Remi Munos.
\newblock Counterfactual credit assignment in model-free reinforcement learning.
\newblock In {\em ICML}, 2021.

\bibitem{cocoa_credit_assignment}
Alexander Meulemans, Simon Schug, Seijin Kobayashi, Nathaniel~D. Daw, and Gregory Wayne.
\newblock Would i have gotten that reward? long-term credit assignment by counterfactual contribution analysis.
\newblock In {\em Advances in Neural Information Processing Systems}, volume~36, 2023.

\bibitem{marti}
Kaiyan Zhang, Kai Tian, Runze Liu, Sihang Zeng, Xuekai Zhu, Guoli Jia, Yuchen Fan, Xingtai Lv, Yuxin Zuo, Che Jiang, Yuru wang, Jianyu Wang, Ermo Hua, Xinwei Long, Junqi Gao, Youbang Sun, Zhiyuan Ma, Ganqu Cui, Ning Ding, Biqing Qi, and Bowen Zhou.
\newblock {MARTI}: A framework for multi-agent {LLM} systems reinforced training and inference.
\newblock In {\em ICLR}, 2026.

\bibitem{drmas}
Lang Feng, Longtao Zheng, Shuo He, Fuxiang Zhang, and Bo~An.
\newblock Dr. {MAS}: Stable reinforcement learning for multi-agent {LLM} systems.
\newblock In {\em Workshop on Multi-Agent Learning and Its Opportunities in the Era of Generative AI}, 2026.

\bibitem{malt}
Sumeet~Ramesh Motwani, Chandler Smith, Rocktim~Jyoti Das, Rafael Rafailov, Philip Torr, Ivan Laptev, Fabio Pizzati, Ronald Clark, and Christian~Schroeder de~Witt.
\newblock {MALT}: Improving reasoning with multi-agent {LLM} training.
\newblock In {\em Second Conference on Language Modeling}, 2025.

\bibitem{naive_assign}
Yujie Zhao, Lanxiang Hu, Yang Wang, Minmin Hou, Hao Zhang, Ke~Ding, and Jishen Zhao.
\newblock Stronger-{MAS}: Multi-agent reinforcement learning for collaborative {LLM}s.
\newblock In {\em ICLR}, 2026.

\bibitem{hong2025multi}
Haoyang Hong, Jiajun Yin, Yuan Wang, Jingnan Liu, Zhe Chen, Ailing Yu, Ji~Li, Zhiling Ye, Hansong Xiao, Yefei Chen, et~al.
\newblock Multi-agent deep research: Training multi-agent systems with m-grpo.
\newblock {\em arXiv preprint arXiv:2511.13288}, 2025.

\bibitem{MAPORL}
Chanwoo Park, Seungju Han, Xingzhi Guo, Asuman~E. Ozdaglar, Kaiqing Zhang, and Joo{-}Kyung Kim.
\newblock Maporl: Multi-agent post-co-training for collaborative large language models with reinforcement learning.
\newblock In {\em {ACL} {(1)}}, pages 30215--30248. Association for Computational Linguistics, 2025.

\bibitem{MARFT}
Junwei Liao, Muning Wen, Jun Wang, and Weinan Zhang.
\newblock {MARFT:} multi-agent reinforcement fine-tuning.
\newblock {\em CoRR}, abs/2504.16129, 2025.

\bibitem{rema}
Ziyu Wan, Yunxiang LI, Xiaoyu Wen, Yan Song, Hanjing Wang, Linyi Yang, Mark Schmidt, Jun Wang, Weinan Zhang, Shuyue Hu, and Ying Wen.
\newblock Re{MA}: Learning to meta-think for {LLM}s with multi-agent reinforcement learning.
\newblock In {\em NeurIPS}, 2026.

\bibitem{TreePO}
Yizhi Li, Qingshui Gu, Zhoufutu Wen, Ziniu Li, Tianshun Xing, Shuyue Guo, Tianyu Zheng, Xin Zhou, Xingwei Qu, Wangchunshu Zhou, Zheng Zhang, Wei Shen, Qian Liu, Chenghua Lin, Jian Yang, Ge~Zhang, and Wenhao Huang.
\newblock Treepo: Bridging the gap of policy optimization and efficacy and inference efficiency with heuristic tree-based modeling.
\newblock {\em CoRR}, abs/2508.17445, 2025.

\bibitem{TreeGRPO}
Yuxiang Ji, Ziyu Ma, Yong Wang, Guanhua Chen, Xiangxiang Chu, and Liaoni Wu.
\newblock Tree search for {LLM} agent reinforcement learning.
\newblock {\em CoRR}, abs/2509.21240, 2025.

\bibitem{treerl}
Zhenyu Hou, Ziniu Hu, Yujiang Li, Rui Lu, Jie Tang, and Yuxiao Dong.
\newblock {T}ree{RL}: {LLM} reinforcement learning with on-policy tree search.
\newblock In Wanxiang Che, Joyce Nabende, Ekaterina Shutova, and Mohammad~Taher Pilehvar, editors, {\em Proceedings of the 63rd Annual Meeting of the Association for Computational Linguistics (Volume 1: Long Papers)}, pages 12355--12369, Vienna, Austria, July 2025. Association for Computational Linguistics.

\bibitem{agentPRM}
Yu~Xia, Jingru Fan, Weize Chen, Siyu Yan, Xin Cong, Zhong Zhang, Yaxi Lu, Yankai Lin, Zhiyuan Liu, and Maosong Sun.
\newblock Agentrm: Enhancing agent generalization with reward modeling.
\newblock In {\em {ACL} {(1)}}, pages 19277--19290. Association for Computational Linguistics, 2025.

\bibitem{DeepSeekMath}
Zhihong Shao, Peiyi Wang, Qihao Zhu, Runxin Xu, Junxiao Song, Mingchuan Zhang, Y.~K. Li, Y.~Wu, and Daya Guo.
\newblock Deepseekmath: Pushing the limits of mathematical reasoning in open language models.
\newblock {\em CoRR}, abs/2402.03300, 2024.

\bibitem{RAG}
Patrick Lewis, Ethan Perez, Aleksandra Piktus, Fabio Petroni, Vladimir Karpukhin, Naman Goyal, Heinrich K{\"{u}}ttler, Mike Lewis, Wen{-}tau Yih, Tim Rockt{\"{a}}schel, Sebastian Riedel, and Douwe Kiela.
\newblock Retrieval-augmented generation for knowledge-intensive {NLP} tasks.
\newblock In {\em NeurIPS}, 2020.

\bibitem{li2025cam}
Rui Li, Zeyu Zhang, Xiaohe Bo, Zihang Tian, Xu~Chen, Quanyu Dai, Zhenhua Dong, and Ruiming Tang.
\newblock {CAM}: A constructivist view of agentic memory for {LLM}-based reading comprehension.
\newblock In {\em NeurIPS}, 2025.

\bibitem{chhikara2025mem0}
Prateek Chhikara, Dev Khant, Saket Aryan, Taranjeet Singh, and Deshraj Yadav.
\newblock Mem0: Building production-ready ai agents with scalable long-term memory.
\newblock {\em arXiv preprint arXiv:2504.19413}, 2025.

\bibitem{personamem}
Bowen Jiang, Zhuoqun Hao, Young~Min Cho, Bryan Li, Yuan Yuan, Sihao Chen, Lyle Ungar, Camillo~Jose Taylor, and Dan Roth.
\newblock Know me, respond to me: Benchmarking {LLM}s for dynamic user profiling and personalized responses at scale.
\newblock In {\em Second Conference on Language Modeling}, 2025.

\bibitem{longmemeval}
Di~Wu, Hongwei Wang, Wenhao Yu, Yuwei Zhang, Kai{-}Wei Chang, and Dong Yu.
\newblock Longmemeval: Benchmarking chat assistants on long-term interactive memory.
\newblock In {\em {ICLR}}. OpenReview.net, 2025.

\bibitem{LOCOMO}
Adyasha Maharana, Dong{-}Ho Lee, Sergey Tulyakov, Mohit Bansal, Francesco Barbieri, and Yuwei Fang.
\newblock Evaluating very long-term conversational memory of {LLM} agents.
\newblock In {\em {ACL} {(1)}}, pages 13851--13870. Association for Computational Linguistics, 2024.

\bibitem{bai2025qwen2}
Shuai Bai, Keqin Chen, Xuejing Liu, Jialin Wang, Wenbin Ge, Sibo Song, Kai Dang, Peng Wang, Shijie Wang, Jun Tang, et~al.
\newblock Qwen2. 5-vl technical report.
\newblock {\em arXiv preprint arXiv:2502.13923}, 2025.

\bibitem{Llama}
Hugo Touvron, Thibaut Lavril, Gautier Izacard, Xavier Martinet, Marie{-}Anne Lachaux, Timoth{\'{e}}e Lacroix, Baptiste Rozi{\`{e}}re, Naman Goyal, Eric Hambro, Faisal Azhar, Aur{\'{e}}lien Rodriguez, Armand Joulin, Edouard Grave, and Guillaume Lample.
\newblock Llama: Open and efficient foundation language models.
\newblock {\em CoRR}, abs/2302.13971, 2023.

\end{thebibliography}

\clearpage
\appendix

\section{Details of Experimental Settings}
\label{app: exp_settings}
\subsection{Datasets}
\label{app: dataset}

\paragraph{PersonaMem.}
PersonaMem \cite{personamem} is a large-scale benchmark for evaluating dynamic user profiling and personalized response generation. It contains more than 180 long-term user--LLM interaction histories and about 6,000 multiple-choice queries, where each history consists of 10, 20, or 60 multi-turn sessions, corresponding to 32k, 128k, and 1M-token settings. Each session contains 15--30 conversation turns and is designed to naturally reveal, reinforce, or update user preferences over time. The subsequent queries cover seven personalization abilities, including recalling user-shared facts, suggesting new ideas, acknowledging recent preferences, tracking preference evolution, revisiting reasons for preference updates, making preference-aligned recommendations, and generalizing preferences to new scenarios.

\paragraph{LongMemEval.}
LongMemEval \cite{longmemeval} is a benchmark for testing the long-term interactive memory of chat assistants. It contains 500 carefully curated questions embedded in scalable, timestamped user--assistant chat histories. The questions are designed to evaluate five core memory abilities: information extraction, multi-session reasoning, temporal reasoning, knowledge updates, and abstention. We use LongMemEval as an external long-memory evaluation setting to test whether the memory system can retrieve and reason over information accumulated across sustained interactions.

\paragraph{LOCOMO.}
LOCOMO \cite{LOCOMO} evaluates very long-term conversational memory for LLM agents. It contains multi-session conversations generated from persona-grounded agents and temporal event graphs, with each conversation spanning hundreds of turns and many sessions. The benchmark covers tasks such as question answering, event summarization, and multimodal dialogue generation. In our experiments, we report LOCOMO question-answering results using F1 and BLEU to evaluate whether the memory system can recover and use long-range conversational evidence.

\paragraph{Data Processing.}
For training data construction, we first identify the session in the full
conversational history that is associated with each query $q$, and denote this
session as the query-specific conversational history $\mathcal{H}$. We then
construct $(\mathcal{H}, q)$ pairs as the basic training examples for memory
construction and retrieval. The position of the target session is provided by
the original benchmark, so our data processing does not require additional
annotation. During evaluation, multiple queries may share the same full
conversational history containing multiple sessions. This setting allows us to
evaluate whether the memory builder and summarizer can organize information from a long history.

\subsection{Detailed Baselines}
\label{app: baselines}
Here, we describe the baseline methods used in our experiments, including methods without explicit memory, prompt-based memory systems, and RL-optimized memory systems.

\paragraph{Base.}
Base directly injects the raw conversational history into the LLM and asks it to answer the query without any explicit memory construction or retrieval module. When the conversation exceeds the model context window, the input is truncated to fit the maximum context length. This baseline measures how well the backbone LLM can exploit long conversational context without an external memory system.

\paragraph{RAG.}
RAG \cite{RAG} retrieves relevant raw dialogue segments before answer generation. We split ultra-long conversational histories into fixed-length chunks of 2,048 tokens, encode them with all-MiniLM-L6-v2, and select chunks through embedding-based similarity matching. The retrieved chunks are then provided to the LLM as evidence for answering the query, serving as a standard retrieval-augmented baseline without learned memory construction.

\paragraph{A-Mem.}
A-Mem \cite{xu2025amem} is an agentic memory system inspired by the Zettelkasten note-taking method. It dynamically organizes memories as interconnected notes and uses dedicated agents for memory construction, precursor memory evolution, and memory retrieval. This design allows newly added memories to update or link to existing memory entries, enabling the memory structure to evolve as the conversation history grows.

\paragraph{Mem0.}
Mem0 \cite{chhikara2025mem0} is a production-oriented long-term memory framework for LLM agents. It extracts salient facts, preferences, and interaction histories from conversations, stores them as persistent memories, and retrieves relevant entries in later sessions. We include Mem0 as a strong prompt-based memory baseline because it explicitly supports memory creation, update, and retrieval across long-running user interactions.

\paragraph{CAM.}
CAM \cite{li2025cam} builds a hierarchical memory system from a constructivist perspective. It uses multiple agents to construct memory nodes through ego-centric disentanglement and node replication, and then aggregates information into higher-level abstractions. We implement CAM on the PersonaMem benchmark to compare our method with a prompt-based multi-agent memory system that emphasizes structured memory construction.

\paragraph{MemoryBank.}
MemoryBank \cite{memorybank} is a long-term memory framework for LLMs that mimics human-like memory behavior. It stores user-related interaction information and introduces a forgetting mechanism based on a forgetting curve, allowing memory strength to decay or be reinforced over time. We implement MemoryBank on PersonaMem as a prompt-based baseline for personalized long-term memory.

\paragraph{Mem1.}
Mem1 \cite{mem1} is an RL-based framework that trains agents to synergize memory and reasoning for long-horizon tasks. Its optimized agents focus on efficient multi-turn information retrieval and reasoning over history, prioritizing direct retrieval from existing conversational records rather than explicit memory construction. We include Mem1 to compare our method with an RL baseline centered on retrieval and reasoning efficiency.

\paragraph{Mem-$\alpha$.}
Mem-$\alpha$ \cite{memory-alpha} is an RL-based memory framework that improves how agents store and access long-term information for downstream reasoning. It optimizes memory-related decisions with reinforcement learning rather than relying only on hand-written prompts. This baseline is used to compare TreeMem with prior RL approaches that improve memory behavior but do not use our tree-based credit assignment.

\paragraph{Memory-R1.}
Memory-R1 \cite{mem-r1} optimizes LLM agents to manage and utilize long-term memories through reinforcement learning. It contains a Memory Manager that learns memory operations such as ADD, UPDATE, and DELETE, and an Answer Agent that uses retrieved memory entries to produce answers. The two agents are optimized independently in stages, making Memory-R1 a strong RL-based baseline for evaluating the benefit of our collaborative multi-agent optimization.

\paragraph{CoMAM.}
CoMAM \cite{CoMAM} is a collaborative multi-agent optimization framework for personalized memory systems. It models memory construction and retrieval agents as a multi-agent system and optimizes them with reinforcement learning to improve final personalized question-answering accuracy. We include CoMAM as a closely related RL-based memory baseline because it also studies coordinated optimization among memory agents, while TreeMem further introduces tree-structured credit assignment for intermediate memory decisions.

\section{Implementation Details}
\label{app: implementation}
All methods are evaluated with the same data splits, input format, and metrics
described in the Section \ref{sec: exp_settings}. We use Qwen2.5-7B-Instruct and
Llama-3.1-8B-Instruct as the backbone LLMs for all compared baselines. All experiments are conducted on a cluster of 8 NVIDIA A100 GPUs.
For each test example, the conversation history is processed in chronological
order. The final prompt contains the user query and the retrieved information,
where the retrieved information is either raw dialogue chunks for retrieval-based baselines
or memory entries for memory-based methods. The answer is then generated under
the same evaluation protocol for all methods.

For the RAG baseline, we follow the setting described above: each long history
is split into chunks of 2,048 tokens, encoded with all-MiniLM-L6-v2, and matched
to the query by embedding similarity. For memory-based baselines, the history is
first converted into persistent memory entries, and the most relevant entries are
retrieved before answer generation. For TreeMem, memory entries are organized as
a tree rather than a flat list. During optimization, the training signal is
propagated along the tree so that both the final answer and the intermediate
memory decisions are optimized. This setting keeps the retrieval and generation
interface comparable across different methods while reflecting the main design
of TreeMem.

For experiments under the 32K, 128K, and 1M context settings, we use the same hyperparameters across all three settings: the batch size is 128, the group size is 8, the maximum model length is 32k, the number of training epochs is 5, and the maximum number of output tokens is 3k.

\subsection{Agent Prompts}
\label{app: prompts}

For reproducibility, we report the prompt templates used by the three TreeMem agents and the answer reward model. The placeholders in braces are replaced with the corresponding conversation history, memory nodes, query, candidate options, or reference answer at runtime.

\vspace{4pt}
\begin{tcolorbox}[colback=black!3!white, colframe=black!70!white, title=Builder Agent Prompt]
\small
\textbf{Role.} You are the TreeMem Builder. Read the chronological conversation history and extract concise, fact-level memory entries for future question answering.

\textbf{Input.} Conversation history: \texttt{\{conversation history\}}.

\textbf{Instructions.} Record durable facts, user preferences, constraints, plans, entities, and updates that may be useful later. Preserve temporal qualifiers when a fact changes over time. Merge duplicate evidence and avoid copying long dialogue spans verbatim. Discard greetings, filler, and unsupported speculation.

\textbf{Output.} Return a list of memory entries. Each entry should contain a short content field, optional time/update information, and brief evidence from the conversation history.
\end{tcolorbox}

\vspace{4pt}
\begin{tcolorbox}[colback=black!3!white, colframe=black!70!white, title=Summarizer Agent Prompt]
\small
\textbf{Role.} You are the TreeMem Summarizer. You receive memory entries from the Builder and organize them into coherent higher-level summaries for the memory system.

\textbf{Input.} Memory entries: \texttt{\{built memory entries\}}.

\textbf{Instructions.} Combine related facts into clear summaries while preserving information needed to answer future questions. Track changes in preferences or important updates explicitly. Do not introduce facts not present in the memory entries. Summaries should reference the relevant underlying entries.
 
\textbf{Output.} Return summaries with main content and pointers to the underlying memory entries they summarize.
\end{tcolorbox}

\vspace{4pt}
\begin{tcolorbox}[colback=black!3!white, colframe=black!70!white, title=Retrieval and Response Agent Prompt]
\small
\textbf{Role.} You are the TreeMem Retrieval agent. Use the memory system to answer the user's query accurately.

\textbf{Input.} Query: \texttt{\{user query\}}. Candidate options, if provided: \texttt{\{options\}}. Memory: \texttt{\{memory entries and summaries\}}.

\textbf{Instructions.} Identify the memory entries and summaries that directly support the query. Prefer specific evidence over broader summaries when they conflict. If the query asks about a changed preference or event, use the most temporally relevant memory. Base the final answer only on retrieved memory and the query.

\textbf{Output.} Put the selected evidence in \texttt{<retrieved\_memory>} and \texttt{</retrieved\_memory>}. Put the final option or free-form answer in \texttt{<final\_answer>} and \texttt{</final\_answer>}.
\end{tcolorbox}

\vspace{4pt}
\begin{tcolorbox}[colback=black!3!white, colframe=black!70!white, title=Answer Reward Model Prompt]
\small
\textbf{Role.} You are the reward model $\mathcal{V}$ used to judge whether a generated answer matches the reference answer for the query.

\textbf{Input.} Query: \texttt{\{user query\}}. Reference answer: \texttt{\{gold answer\}}. Generated answer: \texttt{\{model answer\}}.

\textbf{Instructions.} Compare the generated answer with the reference answer semantically. Treat paraphrases as correct when they express the same choice or fact. Penalize answers that contradict the reference, rely on irrelevant memory, or omit the required fact. Do not reward unsupported extra information unless the required answer is still unambiguous.

\textbf{Output.} Return only \texttt{<score>} followed by a scalar in $[0,1]$ and then \texttt{</score>}.
\end{tcolorbox}

\section{Case Study on Agent-Specific Rewards}
\label{app:agent_specialization}

We report Memory examples from the Summarizer agent in the PersonaMem dataset to illustrate how agent-specific rewards can improve agent quality. The examples are from the Summarizer agent trained under different reward functions. The results demonstrate the superiority of agent-specific rewards.

\vspace{4pt}
\begin{tcolorbox}[colback=red!2!white, colframe=red!65!black, title=Memory from the Summarizer Agent Trained with Unified Final Rewards]
\small
\textbullet~ The user recently had unexpected expenses, struggled to balance their budget, and felt disheartened when new bills arrived. \newline
\textbullet~ They are researching stocks and mutual funds but feel it takes a lot of time. \newline
\textbullet~ They attended a networking event, thought everyone was insincere, and decided to stop attending them. \newline
\textbullet~ They watched personal finance documentaries and liked how visual storytelling simplified complex subjects. \newline
\textbullet~ They quit a negotiation practice group because the intense atmosphere caused severe anxiety and fear of failing. \newline
\textbullet~ They wrote a blog post about their financial challenges, finding the process therapeutic and cathartic. \newline
\textbullet~ They started a personal finance podcast after a friend encouraged them. \newline
\textbullet~ They reduced their volunteer hours for a financial literacy program because they were experiencing burnout. \newline
\textbullet~ They attended a cryptocurrency seminar but got frustrated by the technical jargon and high market volatility. \newline
\textbullet~ They conducted a detailed analysis of their current holdings, looking at past performance and undervalued sectors. \newline
\textbullet~ They found a new couponing method that feels like a scavenger hunt, saves money, and plan to start a savings community. 
\end{tcolorbox}

\vspace{4pt}
\begin{tcolorbox}[colback=orange!3!white, colframe=orange!70!black, title=Memory from the Summarizer Agent Trained with manually designed task-specific Rewards]
\small
\textbf{Interests:} Personal finance, investing, economics, couponing. \newline
\textbf{Dislikes:} Cryptocurrency, networking events, negotiation practice. \newline
\textbf{Activities:} Writes a blog, hosts a podcast, watches documentaries, analyzes their stock portfolio. \newline
\textbf{Current State:} Adjusting budget to save money; trying to balance life and avoid burnout. 
\end{tcolorbox}

\vspace{4pt}
\begin{tcolorbox}[colback=green!3!white, colframe=green!50!black, title=Memory from the Summarizer Agent Trained with Agent-Specific Rewards]
\small
\textbf{User Persona \& Traits:} Highly reflective, intellectually curious, and self-directed. Values authenticity and community building but strictly guards personal boundaries and mental well-being (e.g., limits volunteering to prevent burnout). Analytically driven but risk-averse. \newline
\textbf{Financial Behaviors:} Actively self-corrects spending habits through practical methods (e.g., strategic couponing). Conducts methodical portfolio analyses, focusing on historical volatility and undervalued sectors. \newline
\textbf{Learning Preferences:} \textit{Prefers} visual storytelling (e.g., documentaries), historical context, and low-pressure, self-paced learning. \textit{Avoids} high-pressure performance scenarios (e.g., live negotiation), volatile/jargon-heavy assets (e.g., cryptocurrency), and insincere networking. \newline
\textbf{Content Creation:} Uses writing (blogging) and audio (podcasting) as therapeutic tools to process financial anxiety and aims to build a pragmatic savings community.
\end{tcolorbox}

\section{Limitation}

The main limitation of TreeMem lies in its higher rollout cost compared with a
simpler multi-agent memory system. In a flat multi-agent memory system, each agent usually
makes decisions within a relatively direct interaction process. In contrast,
TreeMem organizes memory operations in a tree structure, which requires the
system to consider more intermediate states and decision paths during
rollout. This design is useful for assigning more fine-grained training signals
to memory construction and retrieval decisions, but it also increases the
computational cost of optimization. Therefore, TreeMem may require more rollout
steps, training time, and GPU resources than simpler multi-agent memory systems,
especially when the conversation history is long or the memory tree becomes
large.

\section{Broader Impact}

This work may have a positive social impact by improving the ability of LLM agents
to maintain and use long-term memory. A more reliable memory system can help
personalized assistants better understand user needs over long interactions,
reduce repeated questions, and provide more consistent support in applications
such as education, personal productivity, and assistive services. TreeMem also
makes memory organization more structured, which may help users and developers
better inspect how useful information is stored and retrieved. We do not identify
any direct negative social impact beyond the general considerations of deploying
LLM-based systems responsibly.


\end{document}